\documentclass[10pt,conference]{IEEEtran}
\IEEEoverridecommandlockouts
\usepackage{booktabs} 
\usepackage{colortbl}
\usepackage{color}
\usepackage{xspace}
\usepackage{url}
\usepackage{balance} 
\usepackage{tabulary}
\usepackage{array}
\usepackage{graphicx}
\usepackage[numbers,sort&compress]{natbib}
\usepackage{subcaption}
\usepackage{listings}
\usepackage{lipsum}
\usepackage{hyperref}
\usepackage{graphicx}
\usepackage{amsmath,amssymb}
\usepackage[T1]{fontenc}
\usepackage{amsmath, amsthm, amssymb}
\usepackage[ansinew]{inputenc}
\usepackage[many]{tcolorbox}
\usepackage{color}
\usepackage{xcolor}
\usepackage{extarrows}
\usepackage{booktabs}
\usepackage{multirow}
\usepackage{colortbl}
\usepackage{tabularx}
\usepackage{enumitem}
\setitemize[0]{leftmargin=15pt}
\graphicspath{{fig/}}
\usepackage[justification=centering]{caption}
\usepackage{pifont}
\usepackage{makecell,multirow,diagbox,rotating}
\usepackage{longtable}
\usepackage{etoolbox} 
\usepackage{listings}

\definecolor{Green}{RGB}{0,180,0}
\newcommand{\revise}[1]{\textcolor{red}{#1}}
\newcommand{\ling}[1]{\textcolor{orange}{#1}}

\newcommand{\xian}[1]{\textcolor{violet}{#1}}

\newcommand{\newtool}{\textsc{ATVHunter}\xspace}

\definecolor{codegreen}{rgb}{0,0.6,0}
\definecolor{codegray}{rgb}{0.5,0.5,0.5}
\definecolor{codepurple}{rgb}{0.58,0,0.82}
\definecolor{backcolour}{rgb}{0.95,0.95,0.92}

\lstdefinestyle{mystyle}{
	backgroundcolor=\color{backcolour},   
	commentstyle=\color{codegreen},
	keywordstyle=\color{magenta},
	numberstyle=\tiny\color{codegray},
	stringstyle=\color{codepurple},
	basicstyle=\ttfamily\footnotesize,
	breakatwhitespace=false,         
	breaklines=true,                 
	captionpos=b,                    
	keepspaces=true,                 
	numbers=left,                    
	numbersep=5pt,                  
	showspaces=false,                
	showstringspaces=false,
	showtabs=false,                  
	tabsize=2
}

\newcommand{\distance}{8pt}
\begin{document}


\title{\newtool: Reliable Version Detection of Third-Party Libraries for Vulnerability Identification in Android Applications}


\author{\IEEEauthorblockN{Xian Zhan\IEEEauthorrefmark{1},
Lingling Fan\IEEEauthorrefmark{2},
Sen Chen\IEEEauthorrefmark{3},
Feng Wu\IEEEauthorrefmark{4},
Tianming Liu\IEEEauthorrefmark{5},
Xiapu Luo\IEEEauthorrefmark{1}, 
Yang Liu\IEEEauthorrefmark{4}
}
\IEEEauthorblockA{\IEEEauthorrefmark{1}Department of Computing,
The Hong Kong Polytechnic University, Hong Kong, China}
\IEEEauthorblockA{\IEEEauthorrefmark{2}College of Cyber Science, Nankai Univerisity, China}
\IEEEauthorblockA{\IEEEauthorrefmark{3}College of Intelligence and Computing, Tianjin University, China}
\IEEEauthorblockA{\IEEEauthorrefmark{4}School of Computer Science and Engineering, Nanyang Technological University, Singapore}
\IEEEauthorblockA{\IEEEauthorrefmark{5}Faculty of Information Technology,
Monash University, Australia}
}

\maketitle

\begin{abstract}
%

Third-party libraries (TPLs) as essential parts in the mobile ecosystem have become one of the most significant contributors to the huge success of Android, which facilitate the fast development of Android applications. {Detecting TPLs in Android apps is also important for downstream tasks, such as malware and repackaged apps identification. To identify in-app TPLs, we need to solve several challenges, such as TPL dependency, code obfuscation, precise version representation. Unfortunately, existing TPL detection tools have been proved that they have not 
{solved} these challenges very well, let alone specify the exact TPL versions.}
%
%
%

To this end, we propose a system, named \newtool, which can pinpoint the precise vulnerable in-app TPL versions and provide detailed information about the vulnerabilities and TPLs.
We propose a two-phase detection approach to identify specific TPL versions. Specifically, we extract the Control Flow Graphs as the coarse-grained feature to match potential TPLs in the pre-defined TPL database, and then extract opcode in each basic block of CFG as the fine-grained feature to identify the exact TPL versions.
We build a comprehensive TPL database (189,545 unique TPLs with 3,006,676 versions) as the reference database.
Meanwhile, to identify the vulnerable in-app TPL versions, we also construct a comprehensive and known vulnerable TPL database containing 1,180 CVEs 
and 224 security bugs. 
Experimental results show \newtool outperforms state-of-the-art TPL detection tools, achieving 90.55\% precision and 88.79\% recall with 
high efficiency,
and is also resilient to widely-used obfuscation techniques and scalable for large-scale TPL detection. 
Furthermore, to investigate the ecosystem of the vulnerable TPLs used by apps, we exploit \newtool to conduct a large-scale analysis on 104,446 apps and find that 9,050 apps include vulnerable TPL versions with 53,337 vulnerabilities and 7,480 security bugs, most of which are with high risks and are not recognized by app developers.

\end{abstract}

\section{Introduction}
\label{sec:Introduction}


Nowadays, over 3 million Android applications (apps) are available in the official Google Play Store~\cite{statista}.
One reason contributing to the huge success of Android could be the massive presence of third-party libraries (TPLs) that provide reusable functionalities that can be leveraged by developers to facilitate the development of Android apps (to avoid reinventing the wheels).
%
 However, extensive TPL usage attracts attackers to exploit the vulnerabilities or inject backdoors in the popular TPLs, which poses severe security threats to app users~\cite{tang2019large,chen2018automated,chen2019gui}. Previous research~\cite{libscout2016ccs,yasumatsu2019CODASPY} pointed out that many apps contain vulnerable TPLs, and some of them have been reported with severe vulnerabilities (e.g., Facebook SDK) that can be exploited by adversaries~\cite{libpecker2018,li2017automatically}.
Attackers can exploit the vulnerabilities in some Ad libraries (e.g., Airpush~\cite{AirPush}, MoPub~\cite{moPub}) to get privacy-sensitive information from the infected devices~\cite{ASIP}.
Even worse, various TPLs are scattered in different apps but the information of TPL components in apps is not transparent. Many developers may not know how many and which TPLs are used in their apps, due to many direct and transitive dependencies. Additionally, about 78\% of the vulnerabilities are detected in indirect dependencies, making the potential risks hard to spot~\cite{SCA_78}.
Thus, vulnerable TPL identification has become an urgent and high-demand task and TPL version detection has become a standard industry product named as Software Composition Analysis (SCA)~\cite{SCA_78,SCA}.
%

Existing TPL detection techniques use either clustering-based methods (e.g., LibRadar~\cite{LibRadar2016ICSE}, LibD~\cite{LibD2017ICSE,LibD22018TSE}) or similarity comparison methods (e.g., LibID~\cite{LibID2019issta}, LibScout~\cite{libscout2016ccs}) to identify TPLs used by the apps. However, according to our analysis and previous study~\cite{libdetect2020ASE}, we conclude the following deficiencies in existing approaches: 
1) \textbf{Low recall.} Clustering-based methods only can identity commonly-used TPLs and may miss some niche and new TPLs, whose recall depends on the number of input apps and the reuse rate of TPLs. Besides, 
the code similarity of different versions and TPL could be various, which makes 
{it difficult to choose appropriate} parameters of the clustering algorithm to perfectly distinguish different TPLs or even versions. 
Verifying the clustering results is also labor-intensive and error-prone. Similarity comparison methods construct a predefined TPL database as the reference database. However, current published size of TPL database is far smaller than the number of TPLs in the actual market thus 
cannot be used to identify a complete set of in-app TPLs.
Apart from that, existing techniques more or less depend on the package structure, especially using package structure to construct the in-app library candidates. Whereas, the package structure/name of the same TPL in different versions could be mutant or easily obfuscated. Therefore, using packages as a supplementary feature to generate TPL signatures is also unreliable~\cite{libdetect2020ASE}.
2) \textbf{Inability of precise version identification.} To find the vulnerabilities of the in-app TPLs, we need to precisely pinpoint the exact TPL versions because not all TPL versions are vulnerable. Even though there are many TPL detection tools, none of them can meet our requirements. AdDetect~\cite{AdDetect2014ISSNIP} just can distinguish the ad and non-ad libraries. ORLIS~\cite{ORLIS2018MOBILESoft} just 
provides the matching class. Clustering-based tools (e.g., LibRadar~\cite{LibRadar2016ICSE}, LibD~\cite{LibD2017ICSE,LibD22018TSE}) do not claim they can pinpoint the exact TPL versions. Besides, current tools~\cite{LibID2019issta,libscout2016ccs,OSSPolice2017CCS,libpecker2018} usually reported many false positives at version-level identification~\cite{libdetect2020ASE}. 
Thus, existing tools are not suitable for vulnerable TPL detection.

Apart from the aforementioned weaknesses of existing tools, we still face some challenges in this research direction:
1) \textbf{Lack of vulnerable TPL version dataset.}
To enable vulnerable TPL version (TPL-V) identification, we need a comprehensive set of known vulnerable TPL-Vs. Ideally, for each vulnerable TPL, it includes TPL names, versions, types, vulnerability severity, etc. However, to the best of our knowledge, no such dataset is publicly available.
2) \textbf{Precise version representation.} We need to distinguish TPLs at version level, however, it is challenging to extract appropriate code features to represent different versions of the same TPL, especially when the code difference of different versions is tiny.
3) \textbf{Interference from code obfuscation.} Many code obfuscation tools (e.g., DashO~\cite{DashO}, Proguard~\cite{Proguard}, and Allatori~\cite{Allatori}) can be used to obfuscate apps and TPLs. For example, 
dead code removal can delete the code without invocation by host apps. These techniques can change the code similarity between in-app TPLs and the original TPLs. Undoubtedly, obfuscation techniques increase the difficulty of TPL identification.

To fill aforementioned research gap, we propose a system, named \newtool (\underline{A}ndroid in-app \underline{T}hird-party library \underline{V}ulnerability \underline{Hunter}), which is an obfuscation-resilient TPL-V detection tool and can report detailed information about vulnerabilities of in-app TPLs. \newtool first uses class dependency relations to split the independent candidate TPL modules from the host app and adopts a two-phase strategy to identify in-app TPLs. It extracts CFGs as the coarse-grained features to locate the potential TPLs in the feature database to achieve high efficiency. It then extracts the opcode sequence in each basic block of CFG as the fine-grained feature to identify the precise version by employing the similarity comparison method. To ensure the recall, we constructed our TPL feature database by collecting  comprehensive and large-scale Java libraries from the maven repository~\cite{maven}. 
We use the fuzzy hash method to generate the signature, which can alleviate the effects from code obfuscation. Compared with previous methods, \newtool does not depend on the package structure.
The main contributions of this work are as follows:
%

\vspace{-1ex}
\begin{itemize}
\item \textbf{An effective TPL version detection tool.}
We propose \newtool, an obfuscation-resilient  TPL-V detection tool with high accuracy that can
find vulnerable in-app TPL-Vs and provide detailed vulnerabilities and components
reports. With the help of our industry collaborator,  \newtool was integrated as a branch of an online service\footnote{\url{https://scantist.io/}} {to help users identify vulnerable Android TPLs.}

    
\item \textbf{Comprehensive datasets.}
We have constructed a comprehensive and large-scale TPL feature database at present, which includes 189,545 TPLs with corresponding 3,006,676 versions to identify in-app TPLs.
We are the first to construct a comprehensive vulnerable TPL-V database for Android apps, including 1,180 CVEs from 957 TPLs with 38,243 vulnerable versions and 224 security bugs from 152 open-source TPLs with 4,533 affected versions.





\item \textbf{Thorough comparisons.} We conduct systematic and thorough comparisons between \newtool and the state-of-the-art tools from different perspectives. The evaluation result demonstrates \newtool is resilient to widely-used obfuscation techniques and outperforms {the} state-of-the-art TPL-V detection tools, achieving high precision (90.55\%) and recall (88.79\%) at version-level identification.
We published the related dataset on our website~\cite{atvhunter}.

\item \textbf{Large-scale analysis.} We leverage \newtool to conduct a large-scale study on 73,110 apps using TPLs and find 9,050
apps contain 10,616 vulnerable TPLs. These vulnerable TPLs include 53,337 known vulnerabilities and 7,480 security bugs. Most of them use TPLs containing severe vulnerabilities. 


\end{itemize}

\section{Related Work} 
\label{sec:relatedwork}

\noindent \textbf{Library Detection.}
AdDetect~\cite{AdDetect2014ISSNIP} and PEDAL~\cite{liu2015efficient} use features such as permissions and APIs to train a classifier to distinguish ad libraries and non-ad libraries. Whereas, these studies fail to identify other types of libraries, such as development aids, UI plugins. 
Currently, there are three TPL detection tools based on the clustering algorithms., i.e., LibRadar, LibD, and LibExtractor. {LibRadar}~\cite{LibRadar2016ICSE} extracts the Android API calls, the total number of API calls and total kinds of API calls as the code features and it chooses the multi-level clustering method to identify potential TPLs. {LibD}~\cite{LibD2017ICSE,LibD22018TSE} extracts the opcode in each CFG block as the code feature. LibExtractor~\cite{LibExtractor2020wisec} exploits the clustering-based method to find potential malicious libraries. 
In general, clustering-based approaches have three common weaknesses: 1) they require a considerable number of apps as input to generate enough TPL signatures. It is also difficult to find emerging or niche TPLs. It also can import some impurities. For instance, if an app is repackaged many times, clustering methods may consider the repackaged host app as a TPL. 2) clustering-based methods may find incomplete TPLs. Some TPLs also depend on other TPLs, but clustering method could separate them into several parts. 3) The above clustering-based approaches more or less rely on package names and package structures, which can be easily obfuscated by existing obfuscators~\cite{Proguard, DashO, Allatori}. LibD claims it is resilient to package name obfuscation and package structure mutation, but package flattening technique can remove the whole package structure and change the internal package structure.
{LibSift}~\cite{LibSift2016soh} constructs the package dependency graph (PDG) to split independent TPL candidates. LibSift does not identify specific libraries, only decouples TPLs into different parts from the host app.
Han et al.~\cite{identifyads2018WPC} aim to measure the behavior differences by comparing benign TPLs and malicious TPLs. It extracts the opcode and Android type tags as features and hashes all feature in each method, and then compare it with the ground-truth libraries to identify the libraries.
{LibScout}~\cite{libscout2016ccs} is a similarity-based library detection tool, which uses the Merkle Tree~\cite{Merkletree} to generate each library instance signature. LibScout chooses the fuzzy method as code feature which changes the non-system identifiers (in the method signature) by using placeholder ``X''. 
{ORLIS}~\cite{ORLIS2018MOBILESoft} uses the same code feature of LibScout~\cite{libscout2016ccs} but different feature generation approach.
LibScout and ORLIS can be resilient to identifier renaming. Whereas, the code feature of LibScout is too coarse, which affects the detection performance.
Besides, ORLIS can only provide the matched class to users, which is not user-friendly. Thus, {they} are not good choices for off-the-shelf TPL detection.
{LibPecker}~\cite{libpecker2018} is also a matching-based library identification tool, which exploits the class dependency as the code features and hashes it as the fingerprint to find TPLs. LibPecker then uses the Fuzzy Class matching method to compare it with the libraries in the database. However, the comparison process is time-consuming. Moreover, LibPecker also assumes the package hierarchy is not change when the TPL is imported into an app, 
{which} will affect the recall.
LibID~\cite{LibID2019issta} is also a TPL version detection tool, but it chooses dex2jar~\cite{dex2jar} as the decompile tool. The reverse-engineering capability of dex2jar directly limits the  detection ability of LibID. More details are clarified in \S~\ref{sec:evaluation}.

\begin{figure*}[t]
	\centering
	\includegraphics[width=\textwidth]{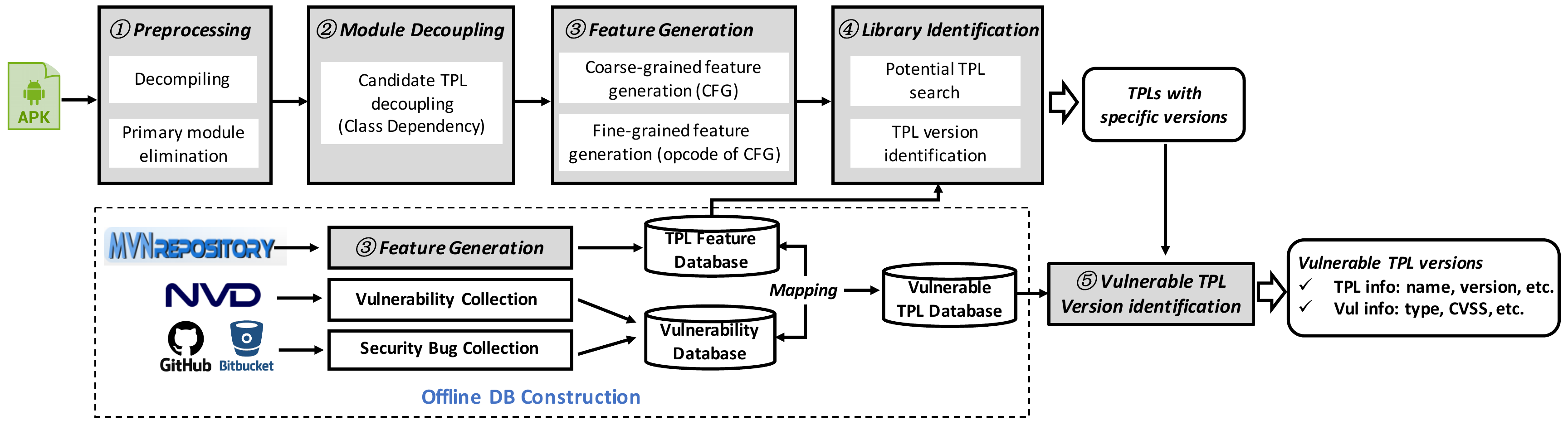}
	\caption{Workflow of \newtool}
	\label{fig:tool_arch}
\end{figure*}

\noindent\textbf{Vulnerable TPL/App Identification.}
{Yasumatsu et al.~\cite{yasumatsu2019CODASPY} attempt to understand how app developers response to the update of TPLs. They studied vulnerable versions of seven TPLs and corresponding apps. By comparing the evolution time between different TPL-Vs and apps versions, they measured the reaction of app developers to these vulnerable TPL versions. 
The number of vulnerable TPL is too small in their dataset, which cannot show the full picture of the infected apps and vulnerable TPLs.}
OSSPolice~\cite{OSSPolice2017CCS} is an automated tool for identifying free software license violations and vulnerable versions of open-source third-party libraries, including both native libraries and Java libraries.
It extracts the fuzzy method signature as the library feature and function centroid~\cite{MassVet2015chen} as {the} version feature to identify TPL-Vs.
However, generating centroid is substantial in terms of resource consumption.
%

\section{Architecture}
\label{sec:approach}

We design a system, \newtool, which takes an Android app as input, and automatically identify the used vulnerable TPL-Vs (if any) according to the constructed database.
Fig.~\ref{fig:tool_arch} shows the system design which is divided into two parts: (1) \textit{TPL-V detection}, which identifies the specific versions of TPLs used by apps; and 
{(2) \textit{vulnerable TPL-V identification}, which can identify the vulnerable in-app TPL-Vs based on our collected known vulnerabilities from NVD~\cite{NVD} and Github~\cite{github}.
Based on the database, we also conduct a large-scale study to assess the ecosystem of Android apps in terms of the usage of vulnerable TPLs.} 
Details are introduced as follows.

\vspace{-0.5ex}
\subsection{TPL Detection}

The TPL detection part of \newtool includes four key phases:
(1) \textit{Preprocessing},
(2) \textit{Module decoupling},
(3) \textit{Feature generation},
and (4) \textit{TPL identification}.

\subsubsection{Preprocessing}

\newtool primarily conducts two tasks in this phase. The first task is to decompile the input app and transform the bytecode into appropriate intermediate representations (IRs). The second task is to find the primary module in the app and delete it to eliminate the interference from the host app. If an app includes TPLs, we call the code of the host app as the ``primary'' module and the in-app TPLs constitute the ``non-primary'' module. 
\newtool first parses the AndroidManifest.xml file and gets the host app packages. Sometimes, the code of the host app may belong to several different namespace, therefore, we need to extract the app packages, application namespace and the package namespace including the Main Activity (i.e., the launcher Activity) and delete these files under the host namespace. {
However, this approach also has following side effects: 1) part of host code suffers from the package flattening or renaming obfuscation and cannot be delete. 2) part of host code cannot be delete due to special package name. 3) the host app and TPLs have the same package namespace, the method may delete these TPLs, leading to false negatives. As for the case 1) \& 2), if the host code and TPLs have no dependencies, it will not affect the accuracy of TPL identification. If the undeleted
 host parts include the TPLs, we can eliminate the interference in the comparison stage. 
}

\subsubsection{Module Decoupling}
\begin{figure}[t]
	\centering
	\includegraphics[width=0.48\linewidth]{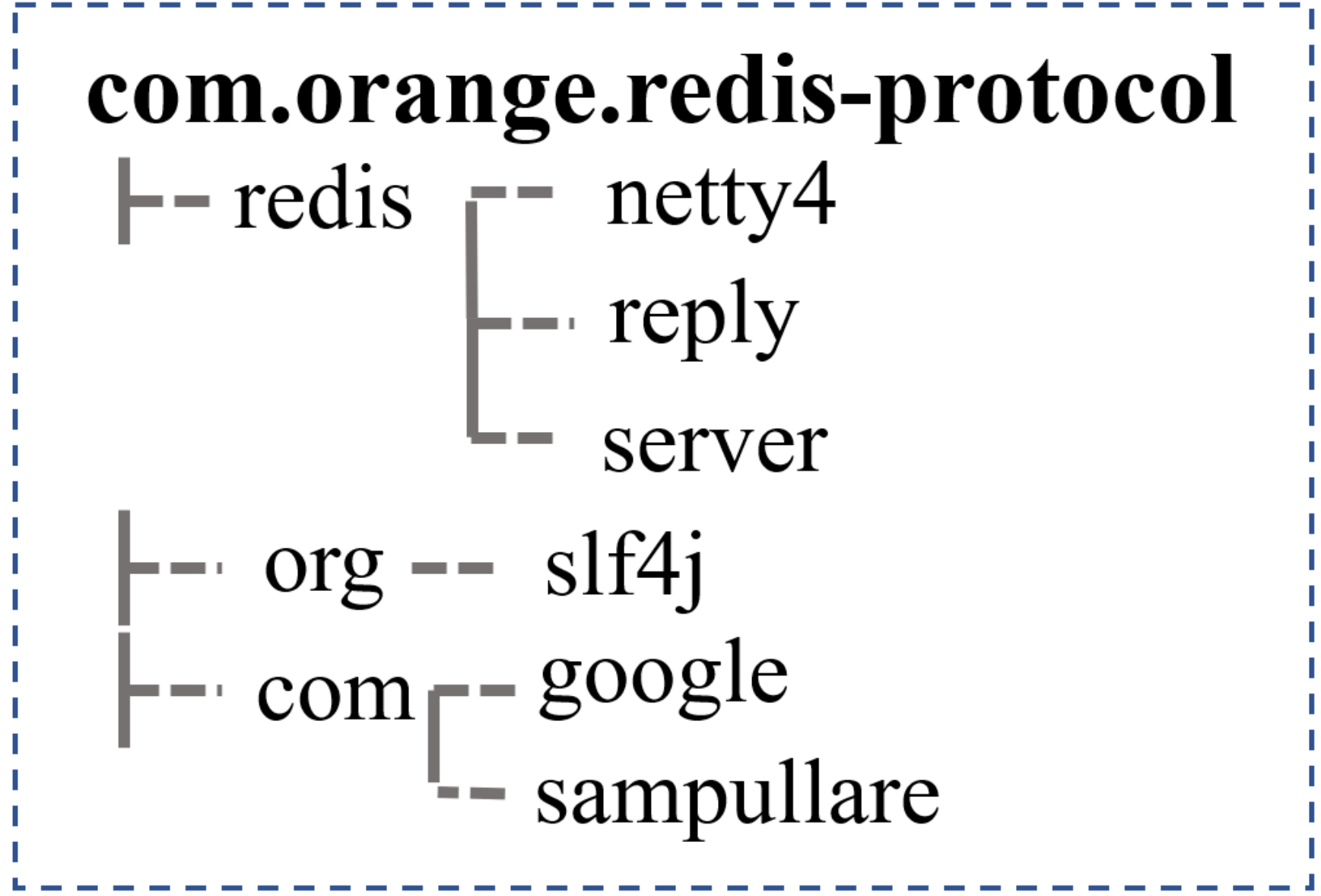}\\
	\caption{An example of a TPL's package structure}
	\label{fig:TPL}
	\vspace{-1ex}
\end{figure}

The purpose of module decoupling is to split up the non-primary module of an app into different independent library candidates.
Previous research adopts different features for module decoupling such as package structure, homogeny graph~\cite{LibD2017ICSE}, and package dependency graph (PDG), however, they more or less depend on the package structure of apps. {Using the package name or the independent package structure to split the in-app TPLs is error-prone, which has two obvious disadvantages: 1) low resiliency to package flattening~\cite{pkg_flt}; 2) inaccurate TPL instance construction. 
{There} are many different TPLs sharing the same root package. For instance, ``com.android.support.appcompat-v7''~\cite{appcompat-V7} and  ``com.android.support.design''~\cite{sup-design} are two different TPLs but the share the same root package \textit{com/android/support}. Besides, one TPL may has multiple parallel package structures, as can be seen an example in Fig.~\ref{fig:TPL}, this TPL\cite{netty4} depends on other TPLs to build itself and developer deploy the ``Fat'' jar mode to package this project. The host TPL with all invoked TPLs constitutes a complete TPL. TPL dependencies are very common, about 47.3\% of Android TPLs in maven repository depend on others based on our rough statistics.}
%
%
To overcome it, we adopt the Class Dependency Graph (CDG) as the features to split up the TPL candidates {because CDG does not depend on the package structure, it is resilient to package flattening}. The class dependency relationship includes: 1) class inheritance, we do not consider the interface relationship because it can be 
{deleted} in obfuscation, 2) method call relationship, and 3) field reference relationship. We use CDGs to find all the related class files, and each CDG will be considered as a TPL candidate in general situation. Using CDGs can avoid the aforementioned situations and package mutation and also be resilient to package flattening.

In \newtool, we use similarity-based method to identify TPL-Vs, we generate the TPL feature database by using the complete TPL files that we downloaded from the maven repository. Therefore, we need to pay attention the packaging techniques of Java projects.
To facilitate maintenance, most developers usually adopt the ``skinny'' mode to package a TPL, which means the released version only contains the code by TPL developers without any dependency TPLs. The dependency TPLs will be loaded during compilation. To solve this situation, we crawl the meta-data of each {TPL} and record their dependency TPLs and packaging technique~\cite{uber} by reading the ``pom.xml'' file. If the ``pom.xml'' claims ``jar-with-dependencies'', it means it includes all dependency TPLs, otherwise, it just 
{includes} the host TPL code. 
If we find a jar {which} is {a} skinny one, 
we also need to split their dependency TPLs by using their package namespace so that we can match the correct version in TPL database.

\subsubsection{Feature Generation} 

After splitting the candidate libraries, we then aim to extract features and generate the fingerprint (a.k.a., signature) to represent each TPL file. To ensure scalability and accuracy, we choose two granularity features. The coarse-grained feature is used to help us quickly locate the potential TPLs in the database. The fine-grained feature is used to help us identify the TPL-V precisely.
(1) For coarse-grained features, we choose to extract the Control Flow Graph (CFG) to represent the TPL 
since CFG is relatively stable~\cite{CFO2016}. 
CFG also keeps the semantic information that ensures the accuracy to some extent~\cite{chen14ICSE}. 
(2) For fine-grained features, we extract the opcode in each basic block of CFG as the feature for exact version identification.

\noindent \textbf{Coarse-grained Feature Extraction.}
We first extract the CFG 
for each method in the candidate TPLs, and traverse the CFG to assign each node a unique serial number (starting from 0) according to the execution order. 
For a branch node with sequence number $n$, its child with more outgoing edges will be given sequence number $n+1$ and the other child is given $n+2$. If two child nodes have the same outgoing edges, we will give $n+1$ to the child node with more statements in the basic block.
We then convert the CFGs into signatures based on the assigned serial numbers of each node to represent each unique TPL, in the form of \texttt{\small [node count, edge adjacency list]}, 
where the adjacency list is represented as: \texttt{[$parent_1$ -> ($child_1$,$child_2$,\dots), $parent_2$ -> \dots]}. 
We then hash the adjacency list of CFG as a method signature. To improve the search efficiency, we sort these hash values in ascending order and then hash the concatenate values as one of the coarse-grained TPL features (T1). Meanwhile, we also keep the series of CFG signatures in our database to represented each TPL in feature database.

\begin{figure}[t]
	\centering
	\includegraphics[width=0.93\linewidth]{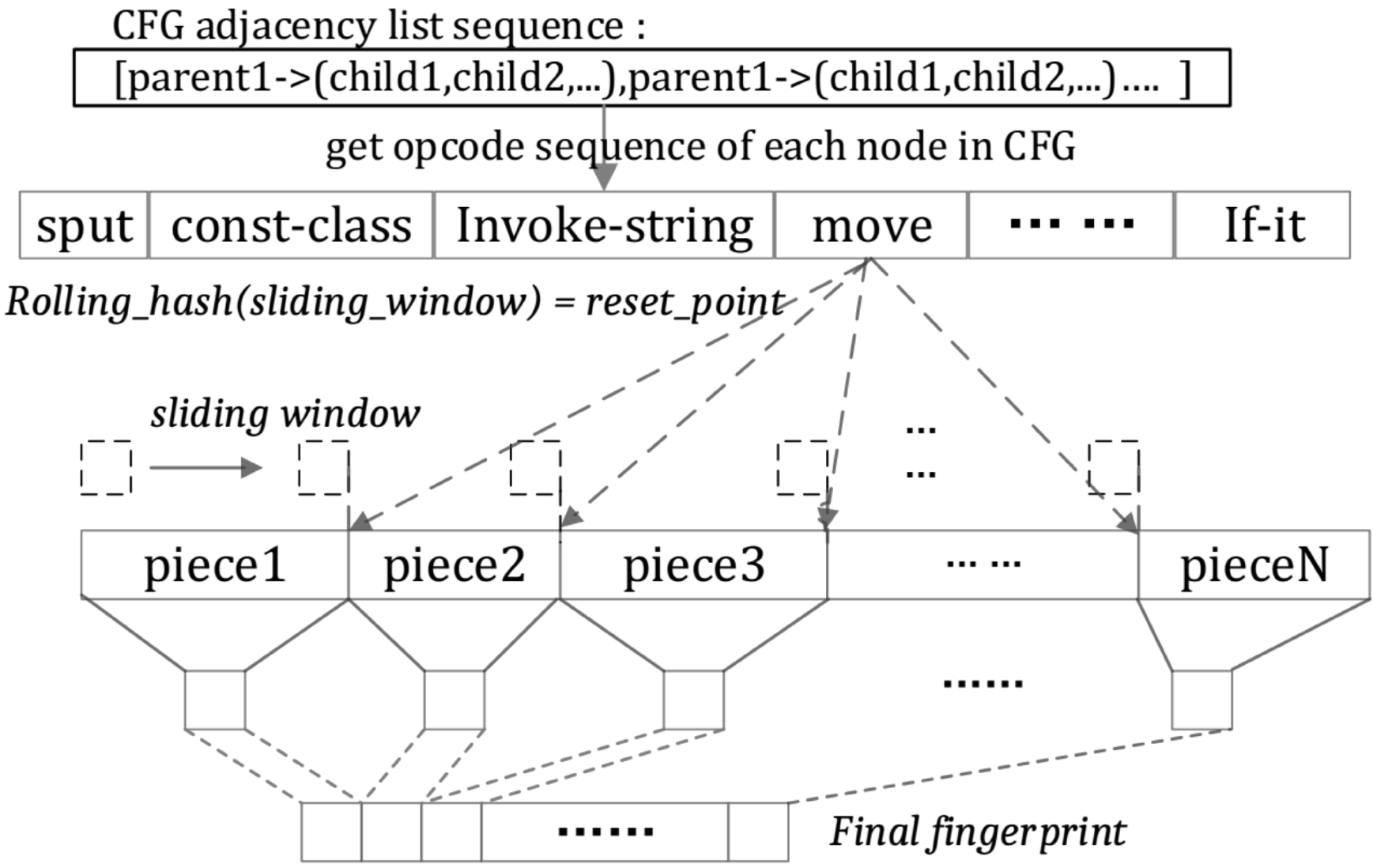}\\
	\caption{Fuzzy hashing for method feature generation as the version feature}
	\label{fig:edit_distance}
\end{figure}

    \noindent \textbf{Fine-grained Feature Extraction.}
Based on our analysis, we find the code similarity of different versions for the same TPL could be diverse, which can range from about 0\% to nearly 100\%. 
The coarse-grained features (i.e., CFG) are likely to generate the same signature of different versions that have minor changes such as insert/delete/modify a statement in a basic block.
Therefore, we propose finer-grained features, i.e., opcode in each basic block of CFG, to represent each version file. 
However, extracting more fine-grained features will increase more computational complexity and cost of the computing resources.
To ensure the scalability of \newtool, a common way to achieve that is through hashing~\cite{DroidMOSS12CODASPY}. However, hash-based method has an obvious drawback to determine whether two objects (e.g., TPLs, methods) are similar because a minor modification can lead to a dramatic change of the hash value. Thus, we adopt the fuzzy hashing technique~\cite{fuzzyhash} instead of the traditional hash algorithm to generate the code signature for each method. 
%
{Fig.~\ref{fig:edit_distance} shows the feature generation process for TPL-Vs.} 
Specifically, we first 
extract all the opcode sequences inside each basic block and concatenate them together.
We do not consider the operands (e.g., identifier names or hard-coded URLs) that 
are not robust for some simple obfuscation techniques such as renaming obfuscation and string encryption techniques~\cite{DashO_obfuscation,DroidMOSS12CODASPY}.
We then concatenate all opcode sequences of each basic block according to the adjacency list of CFG. In this step, our method is somewhat similar to LibD~\cite{LibD2017ICSE} with respect to the code feature. We also adopt the opcode in each basic block of CFG as the code feature. However, we also have many differences. LibD uses a package-level hash value as the final signature and {uses the} clustering algorithm to detect TPLs.
While in \newtool, to defend against code obfuscation or TPL customization~\cite{libpecker2018}, we use the fuzzy hash on each method-level feature and similarity comparison to find similar methods. We first use a slide window (a.k.a., rolling hash~\cite{fuzzyhash}) to cut the opcode sequence into small pieces. Each piece has an independent contribution to the final fingerprint. If one part of the feature changes due to code obfuscation, it would not cause a big difference to the final fingerprint.
We then hash each piece and combine them as the final fine-grained fingerprint of each method. The fingerprints of all methods in a version to represent a TPL-V.


\noindent\textbf{TPL Database Construction.}
We crawled all Java TPLs from Maven Repository~\cite{maven} (189,545 unique TPLs with their 3,006,676 versions) to build our TPL database. We use the above mentioned method to obtain the signature for each TPL. 
For each version of TPLs, we store both coarse-grained and fine-grained features in a MongoDB~\cite{mongoDB} database. The size of the entire database is 300 GB. We spent more than one month to collect all the TPLs and another two months to generate the TPL feature database.

\noindent \subsubsection{Library Identification}
\label{sec:approach:lib_identification}
This step aims to identify the used TPL-Vs in a given app. 
To achieve it efficiently, we propose a two-stage identification method: 1)  potential TPL identification; 2) version identification.

\noindent\textbf{1) Potential TPL Identification.}
Since there are over 3 million TPL files to be compared in our database for each candidate library, to speed up the entire detection process, we search the database in the following order:  a) \textit{Search by package names.}
For each library candidate, we first use its package namespace (if not obfuscated) to narrow down the search space in our database. 
Note that we cannot directly use the package name to determine a TPL, because the same package namespace could include different third-party libraries. For example, the Android support group~\cite{android_support} includes 99 different TPLs. These TPLs have the same group ID ``\textit{com.android.support}'' and 
the same package name prefix ``\textit{android/support/}''. 
If the package name has been obfuscated or a candidate TPL module is without a package name, we move to the next filtering strategy.
Note that, 
even though it is a non-trivial problem to decide the obfuscated package name,
in our work, the package name is only used as supplementary information to speed up the search process. No matter whether a candidate TPL can find a match in the TPL database by using the package names, we still continue to search the TPL database via other features. Thus, we only applied a simple rule to identify the obfuscated apps: if a package name is a hash value or a single letter, we consider it obfuscated.
b) \textit{Search by the number of classes.} We assume two TPLs are unlikely to be the same one if the number of classes within two TPLs has a big difference~\cite{ViewDroid14}. 
If the number of the classes in a TPL only accounts for less than 40\% of that in another TPL in the database, we will not further compare them, which can help us speed up the identification process. 
c) \textit{Search by coarse-grained features.} To speed up, we first search the coarse-grained feature T1 in the TPL database; if we find the same one, ATVHunter will report this TPL and stop the search process. Otherwise, \newtool will 
compare the candidate TPL with TPLs in the database, if all the coarse features are the same, we consider find the TPL and the search process will stop.
If over 70\% of the coarse-grained features are the same (followed by previous research~\cite{DroidMOSS12CODASPY,MassVet2015chen,ViewDroid14,DroidSim2014IFIP}), we consider it as a potential TPL. 
When we find the potential TPL, we will identify the exact version.

\noindent \textbf{2) Version Identification.}
{To identify the specific versions of the used TPLs, we utilize the fine-grained features and calculate the similarity ratio of two TPLs as the evaluation metric. 
To ensure the efficiency, we do not compare these matched methods in previous stage. \newtool can record the same method pair in the previous stage, therefore, we only need to compare less than 30\% of the methods in this phase.
Since some code obfuscation techniques (e.g., junk code insertion) would change the fingerprints of methods, causing two methods that were initially the same to be different. Therefore, we need to compare the method similarity and consider two methods matched only when their method similarity exceeds a threshold. 
Based on the number of matched methods, we then compute the TPL similarity. When the number of matched methods exceeds the threshold, we consider we find the correct TPL with its version.
} 

\noindent $\bullet$ \textit{Method Similarity Comparison.}
We employ \textit{edit distance}~\cite{DroidMOSS12CODASPY,edit_distance} to measure the similarity between two method fingerprints. 
The edit distance of two fingerprints is defined as the number of minimum edit operations (i.e., insertion, deletion, and substitution) that is required to modify one fingerprint to the other.
%
%
Based on the edit distance of two signatures, we compute the Method Similarity Score (MSS) between two methods (i.e., $m_a$ and $m_b$) 
by using the formula:
\vspace{-2mm}
\begin{equation}
   MSS(m_a, m_b) = 1- \frac{d[m_a, m_b]}{max(m,n)} 
   \label{equ:mthd:sim}
\end{equation}
where $m$ and $n$ represent the signature length of two methods and $d[m_a,m_b]$ is the edit distance of two method signatures.
If $MSS$ exceeds a certain threshold $\theta$, we consider the two methods are matched. Based on our experimental result in \S~\ref{sec:evaluation:RQ1}, we choose {$\theta=0.85$} as the threshold. 

\noindent $\bullet$ \textit{TPL Similarity Comparison.}
Based on the number of matched methods, the similarity of two TPLs ($t_1$ and $t_2$) are defined as follows:
\vspace{-1.3ex}
\begin{equation}
TSS(t_{1},t_{2}) = \frac{M_{|t_1\bigcap t_2|}}{M_{|t_{2}|}}
\label{equ:TPL:sim}
\end{equation}
where $t_1$ is a TPL candidate from the test app, $t_2$ is a TPL from the database for comparison.
$M_{|t_{2}|}$ is the number of methods in $t_{2}$. 
$M_{|t_1\bigcap t_2|}$ is the number of matched methods of $t_1$ and $t_2$ which should meet two conditions: (a) $\forall m_i, m_j$, where $m_i$ is a method of $t_1$, $m_j$ is a method of $t_2$, $MSS(m_i, m_j) \geq  \theta$; (b) $\exists m_j$, that $MSS(m_i, m_j) = 1$, that is, we only compare two TPLs that have at least one exactly matched method in order to speed up the identification process.
For a TPL candidate $t_1$, we consider we find a potentially matched TPL-V ($t_2$) in the database when $TSS(t_{1},t_{2})\geq \delta$, $\delta$ is the similarity threshold, and select the TPL-V with the largest similarity score as the final result of $t_1$, providing the identified TPLs with group id, artifact id and version number. We set the threshold $\delta = 0.95$ based on our experimental result in \S~\ref{sec:evaluation:RQ1}.

%

%

\subsection{Vulnerable TPL-V Identification}

We first build a vulnerable TPL-V database, based on which we identify the vulnerable TPL-Vs used by the apps.

\subsubsection{Database Construction}
The vulnerable TPL-V database construction process includes collection of know vulnerabilities in Android TPLs and security bugs from open-source software.

\noindent \textbf{Known TPL Vulnerability Collection.}
{To collect the vulnerable TPL versions, we convert the names of all TPL files (3,006,676 in total) in our feature database into Common Platform Enumeration (CPE) format~\cite{CPE} and exploit \textit{cve-search}~\cite{cve-search}, a professional CVE search tool, to query the vulnerable TPLs from the public CVE (Common Vulnerabilities and Exposures) database by mapping the transformed TPL names. 
In this way, we can get the known vulnerabilities of TPL-Vs and their detailed information, including the CVE id, vulnerability type, description, severity score from Common Vulnerability Scoring System (CVSS)~\cite{cvss}, vulnerable versions, etc. We use \textit{CVSS v3.0} to indicate the severity of the collected vulnerabilities in this paper. 
Finally, we collected 1,180 CVEs from 957 unique TPLs with 38,243 affected versions.}

\noindent \textbf{Security Bug Collection.} 
Since \newtool is able to identify the specific versions of TPLs used by apps, therefore, besides the known vulnerabilities, we also obtain 224 security bugs from Github~\cite{github} and Bitbucket~\cite{bitbucket} owing to the collaboration with our anonymous industrial collaborators.
These bugs come from 152 open-source TPLs with their corresponding 4,533 versions.
All of these security bugs have been cross-validated by the security experts in industry.

\subsubsection{Vulnerable TPL-V Identification}
When \newtool identifies the used TPL-Vs in the app, it will search the vulnerable TPL database to check whether these identified TPL-Vs are vulnerable or not. If \newtool finds the vulnerable TPL-Vs, it will generate a detailed vulnerability report to users.
We believe \newtool can serve as an extension of ASI Program~\cite{ASIP} for Google. The previous research~\cite{yasumatsu2019CODASPY} reported that vulnerabilities listed on ASI program can draw more attention to developers. However, the vulnerabilities are reported by ASI program is limited. Our comprehensive dataset can be a supplement to ASI program. 

\vspace{-0.5ex}
\subsection{Implementation}
\label{sec:implementation}

\newtool is implemented in 2k+ lines of python code.
We employ \textsc{Apktool}~\cite{apktool}, a reverse engineering tool commonly-used by much previous work~\cite{fan2018large,crash2020,chen2019storydroid,fan2018efficiently} to decompile the Android apps and exploit Androguard~\cite{Androguard} to obtain the class dependency relations in order to get the independent TPL candidates.
We then employ \textsc{Soot}~\cite{soot} to generate CFG and also build on \textsc{Soot} to get the opcode sequence in each basic block of a CFG.
We use the ssdeep~\cite{ssdeep} to conduct fuzzy hash algorithm to generate the code feature and employ the edit distance~\cite{edit_distance} algorithm to find the in-app TPLs. Our approach can pinpoint the specific TPL versions.
We maintain a library database containing more than 3 million TPL files and construct a vulnerable TPL database that includes 224 security bugs from open-source Java software 
{on} Github, and 1,180 CVEs from 910 Android TPLs in public CVE databases.

\section{Evaluation}
\label{sec:evaluation}

In this section, we first construct our ground truth and choose appropriate thresholds for MSS and TSS in \S~\ref{sec:evaluation:RQ1}.
Based on the thresholds, we further evaluate \newtool from \textbf{effectiveness (RQ1)}, \textbf{scalability (RQ2)}, and \textbf{the capability of code obfuscation-resilience (RQ3)}.
All the experiments were conducted on a commercial cloud service running Ubuntu 16.04 LTS with 8-core Intel(R) Xeon(R) Gold 6151 processor, CPU @ 3.00GHz and 128G memory.


\subsection{Preparation}
\label{sec:evaluation:RQ1}
\noindent $\bullet$  \textbf{Ground-truth Dataset Construction.} We build this dataset for three primary purposes: 1) verify the effectiveness of \newtool; 2) compare the performance with {the} state-of-the-art tools; 3) release the datasets to the community to promote follow-up research.
{Since it is difficult to know the specific TPL-Vs from commercial apps,} we choose the open-source apps to 
{compare \newtool with} existing tools.

We first collect the latest versions of 500 open-source apps from F-Droid~\cite{F-Droid} that is the largest repository maintaining open-source Android apps. 
We choose open-source apps as subjects since we can get the specific TPL information (including the version) in the configuration files and source code of apps, such a mapping relation between apps and TPLs is used as the ground-truth for performance evaluation. These apps are from 17 different categories with various sizes.
For each app, we manually analyze it and get the in-app TPLs with their specific versions.
According to our analysis, these apps contain the number of TPLs ranging from 2 to 37 and these TPLs also have different functions with diverse sizes. We then download these TPLs with their versions from the Maven repository~\cite{maven}. 
To ensure the evaluation results more reliable, we collect the complete versions of each TPL. We filter 144 apps out due to the incomplete versions of TPLs maintained in the Maven repository. Note that, based on our analysis, we find the previous published datasets have some biases. TPLs from LibScout and LibID are most independent ones, thus, we add some TPLs that depend on other TPLs in our dataset, such as ``Retrofit'' depends on ``Guava'', to reveal the lib identification capability of different tools. 
Finally, we choose 356 apps and 189 unique TPLs with the complete 6,819 version files in these apps as the ground truth. 


\begin{figure}[t]
	\centering
	\begin{subfigure}{0.25\textwidth}
		\centering
		\includegraphics[scale=0.26]{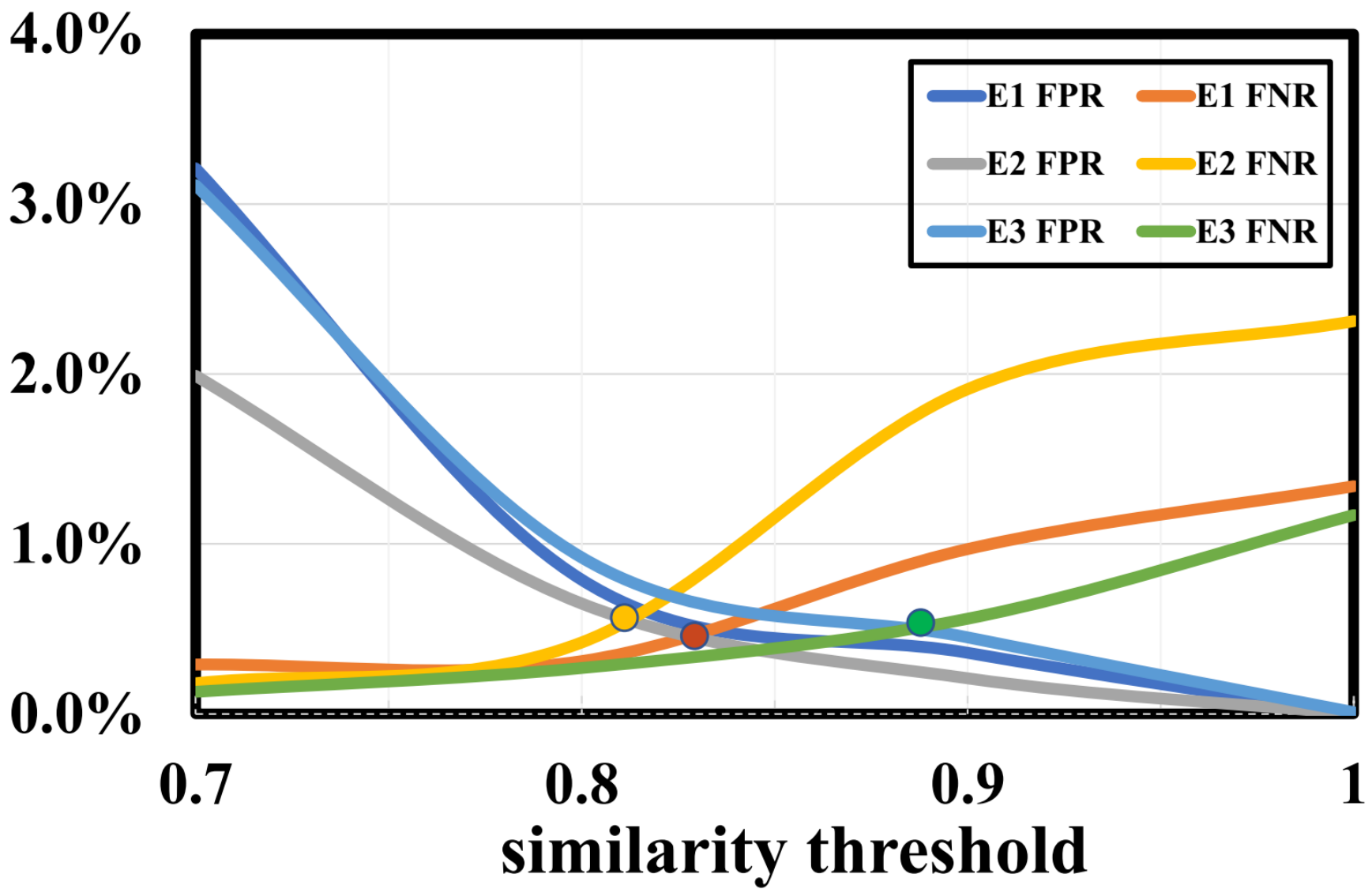}
		\caption{Method-level}
		\label{fig:threshold_method}
	\end{subfigure}%
	\begin{subfigure}{0.24\textwidth}
		\centering
		\includegraphics[scale=0.25]{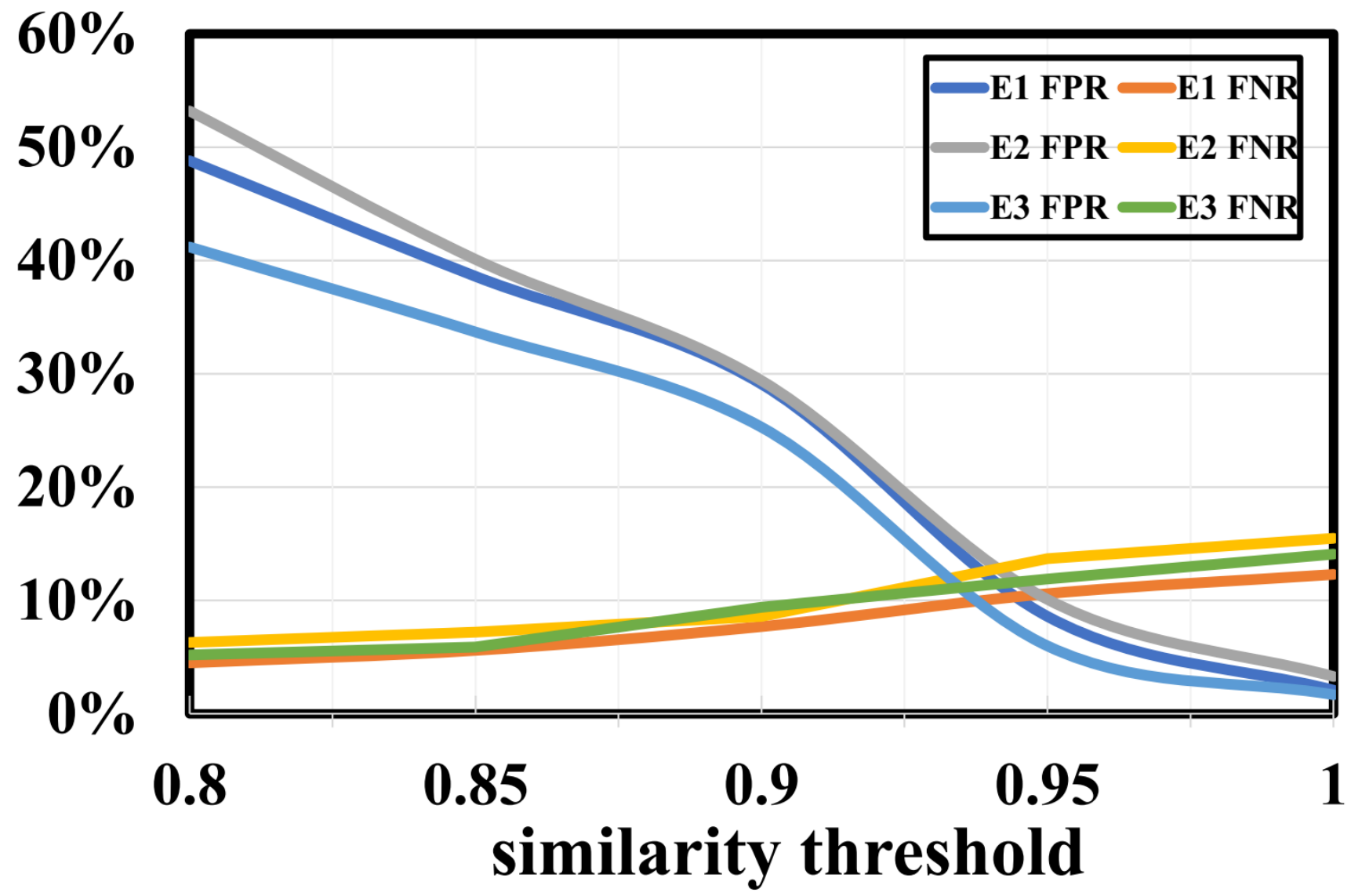}
		\caption{TPL-level}
		\label{fig:threshold_TPL}
	\end{subfigure}

	\caption{Similarity threshold selection}
	\label{fig:Threshold determination}
	\vspace{-1ex}
\end{figure}

\noindent $\bullet$ \textbf{Threshold Selection.} 
%
To avoid bias, we randomly select three groups ($3\times200$) of apps except the aforementioned dataset to decide appropriate thresholds for method similarity score $\theta$, and TPL similarity score $\delta$.
We use method-level false positive rate (FPR) and false negative rate (FNR); and TPL-level FPR and FNR as the metrics to decide the similarity thresholds by varying $\theta$ and $\delta$ from different thresholds. We employ the three groups of apps to implement the same experiment three times and then decide the optimal thresholds.


Fig.~(\ref{fig:threshold_method}) shows the method-level FPR and FNR at different similarity thresholds. 
We can find when the threshold $\theta$ is around 0.85,
both the FPR and FNR are relatively low. Therefore, we choose $\theta = 0.85$ as the MSS threshold where the FPR is less than 1\% and FNR is less than 0.5\%, which can achieve a good trade-off. Fig.~(\ref{fig:threshold_TPL}) shows the TPL-level FPR and FNR
at different thresholds. 
According to the result, we find that when the threshold is gradually close to 0.8, many false positives appear due to the same TPL with different minor-changed versions. When the threshold is close to 1, the number of false negatives increases. 
From Fig.~\ref{fig:Threshold determination}, 
we can find FPR and FNR achieve a good trade-off when the threshold is around 0.95,
we thus choose {0.95} as the threshold $\delta$ of TSS.
In summary, we employ $\theta = 0.85$ and {$\delta = 0.95$} for the following experiments.

\vspace{-1ex}
\subsection{RQ1: Effectiveness Evaluation}
\label{sec:evaluation:RQ2}

\noindent\textbf{Experimental Setup.}  
For the effectiveness evaluation, we compare \newtool with the {state-of-the-art} publicly-available TPL detection tools ({i.e., LibID, LibScout, OSSPoLICE, and LibPecker}) that can specify the used TPL versions by using our ground truth dataset (\S~\ref{sec:evaluation:RQ1}). 
We employ three evaluation metrics, i.e., precision ($\frac{TP}{TP + FP}$), recall ($\frac{TP}{TP + FN}$) and F1 Score ($\frac{2*Precision*Recall}{Precision + Recall}$), to evaluate the detection accuracy at both TPL-level and version-level. 
TPL-level identification indicates the ability to identify the in-app TPLs correctly (without specifying the versions), and version-level identification indicates the ability to find both the correct TPLs and the correct versions.  For example, if a tool reports that it finds  ``okio-2.0.0, okio-2.3.0'' in an app but the ground truth is ``okio-2.4.3'', in this situation, for TPL-level, we consider the tool find the correct TPL; for version-level, we consider there are two false positives and one false negative.

%
%

%


\noindent{\textbf{Result.}} Table~\ref{tbl:accuracy} shows the comparison results of \newtool and other state-of-the-art tools. Considering the overall performance, we can see \newtool outperforms other tools regarding all the metrics; the F1 score of \newtool at library-level and version-level reached 93.43\% and 88.82\%, respectively.
{For library-level identification, we can find that all of them can achieve high precision at TPL-level identification but the performance of recall of current state-of-the-art tools is mediocre. In contrast, the recall of \newtool is 88.79\%, which is far better than others.}
{For version-level identification,
we can find the precision (90.55\%) and recall (87.16\%) of \newtool is much higher than that of other tools. Compared with the library-level precision, we can see the precision of each tool at version-level decreases a lot, which means most of them can identify the TPL but they cannot pinpoint the exact versions.} We elaborate on the reasons for false positives and false negatives of \newtool and other state-of-the-art tools as follows.

%

\noindent \textbf{FP Analysis.}
The reasons for the false positives of \newtool can be concluded in three points: 
(1) \textit{reuse of open-source components.} We find some TPLs are re-developed based on other TPLs, with only small code changes, if their similarity is larger than the defined threshold, \newtool will report the reused ones at the same time, which are false positives.
(2) \textit{Artifact id or group id changes.}
We identify a TPL by using its group id, artifact id and version number. However, we find that some old version TPLs has migrated to the new ones, with their group id or artifact id changed, but their code has little difference.
Take the TPL file ``EventBus'' as an example, ``org.greenrobot:eventbus''~\cite{greenEventBus} is the upgraded version of ``de.greenrobot:eventbus''~\cite{EventBus_old}. The code of these two TPLs have high similarity but with different group ids. 
\newtool matches both of them and considers they are different TPLs. 
(3) \textit{Different versions with high similarity}. 
The other reason for the false positives of \newtool is that some versions of the same TPL have little or no difference in their code.
For example, ``ACRA\_4.8.3'' only modifies a few statements in a method of ``ACRA\_4.8.2'', and \newtool would report the two versions of the TPL at the same time, one of them is regarded as false positives.
In our database, we even find some versions of the same TPL have the same Java code but different resource files, configuration files or native code (C/C++), but this situation does not affect the vulnerable TPL identification process. 

%
As for the false positives of other tools, the code feature of LibScout (i.e., fuzzy method signature) is too coarse, which would make it generate the same signature for different versions if the two versions have minor differences. As the aforementioned example ``ACRA'', all existing tools cannot distinguish the two versions because it generates the same signature for them. Besides, if the methods are very simple, the signatures generated by LibScout and OSSPoLICE would also be the same, which can also lead to false positives. LibPecker depends on the package structure as a supplementary feature to identify different TPLs, they may report a TPL depend on others TPLs several times. For instance, if an app use the Library C that is built on library A and B, if library A and B are also in TPL feature database, LibPecker could report library C as library A and B, leading to false positives.



 %

\begin{table}[t]

\centering
\small
\caption{Library and Version Detection Comparison}
\scalebox{0.8}{
\begin{tabular}{l|lll|lll}
\hline
\multicolumn{1}{c|}{\multirow{2}{*}{\textbf{Tools}}} &
  \multicolumn{3}{c|}{\textbf{Library-level}} &
  \multicolumn{3}{c}{\textbf{Version-level}} \\ \cline{2-7} 
\multicolumn{1}{c|}{} &
  \multicolumn{1}{c}{\textbf{Precision}} &
  \multicolumn{1}{c}{\textbf{Recall}} &
  \multicolumn{1}{c|}{\textbf{F1}} &
  \multicolumn{1}{c}{\textbf{Precision}} &
  \multicolumn{1}{c}{\textbf{Recall}} &
  \multicolumn{1}{c}{\textbf{F1}} \\ \hline
\textbf{ATVHunter} & \textbf{98.58\%} & \textbf{88.79\%} & \textbf{93.43\%} & \textbf{90.55\%} & \textbf{87.16\%} & \textbf{88.82\%} \\
\textbf{LibID} & 98.12\% & 68.45\% & 80.64\% & 68.70\% & 66.42\% & 67.54\% \\
\textbf{LibScout} & 97.10\% & 46.65\% & 63.02\% & 44.82\% & 43.50\% & 44.15\% \\
\textbf{OSSPoLICE} & 97.91\% & 43.39\% & 60.13\% & 88.83\% & 42.25\% & 57.26\% \\
\textbf{LibPecker} & 93.16\% & 57.82\%  & 71.35\% & 60.35\% & 57.67\% & 58.98\% \\ \hline

\hline

\end{tabular}
}
\label{tbl:accuracy}
\end{table}

\noindent\textbf{FN Analysis.}
{\newtool aims to find TPL versions with high precision,}
thus, we sacrificed part of the recall when we select the similarity threshold. The reasons for false negatives of \newtool are as follows:
(1) When compiling an app, developers may take some optimizations to reduce the size of their app. The strategy is that the compiler automatically removes some functions of TPLs that are not called by host apps, which causes the in-app TPLs to be different from the original TPLs, leading to false negatives. 
(2) Some TPLs are integrated into the same package namespace of the host app, which may be deleted at the pre-processing stage, leading to false negatives. For example, some companies and organizations develop their own Ad SDK, whose package name is the same as that of the host app. However, the code under the package structure of the host app is deleted at the pre-processing stage, i.e., the ad library is also deleted without further consideration, causing the false negatives.
(3) Another reason is that some apps use rarely-used open-source TPLs hosted on open-source platforms (e.g., Github or Bitbucket) which are not in our TPL database (with over 3 million TPLs), leading to false negatives. 
For example, the TPLs ``com.github.DASAR.ShiftColorPicker'', ``android-retention-magic-1.2.2'', and ``android-json-rpc-0.3.4'' are developed and hosted on Github, and not in our dataset, therefore, \newtool cannot find this TPL. 
Since other tools also use the similarity comparison method to find in-app TPLs, this situation also may affect their recall.

%
%
%

As for the false negatives of other TPL detection tools, they more or less use the package structure to generate the TPL features. However, the package structure is not stable, which can be easily changed by the package flattening obfuscation. 
{We find the packages structures of many real-world in-app TPLs are more or less obfuscated, and some TPLs are even without any package structure;} current tools cannot handle 
{such cases,} leading to false negatives.
Besides, it is difficult to use the package structure and package name to ensure the TPL candidates, ling{as demonstrated in \S\ref{sec:approach:lib_identification}.} 
Many different TPLs may have the same package name, and one independent package tree could include several TPLs; therefore, existing tools may generate incorrect code features for these TPLs, which also can lead to false negatives. LibID uses Dex2jar~\cite{dex2jar} to decompile apps, it does not always work in all apps, which discounts the recall of LibID. Besides, LibScout and OSSPoLICE are sensitive to CFG structure modification. 
%

Compared with them, our CFG adjacency list is less sensitive to the CFG structure modification. We consider both the syntax and semantic information, and our method adopts the fuzzy hash to generate the TPL fingerprints. Thus, code statements modification can only affect part of the fingerprints, which is more robust to different code obfuscations.
Based on the above analysis, we can find that the strategy of feature selection, extraction, and generation are essential, which can directly affect the performance of the system.

\smallskip
\noindent\fbox{
	\parbox{0.95\linewidth}{
		\textbf{Conclusion:}
		\newtool outperforms state-of-the-art TPL detection tools, achieving 98.58\% precision, 88.79\% recall at library level, and 90.55\% precision, 87.16\% recall at version level. 
	}
}

\subsection{RQ2: Efficiency Evaluation}
\label{sec:evaluation:RQ3}
In this section, we investigate the detection time of \newtool and compare it with state-of-the-art tools to verify its efficiency. 
%
{We compare the detection time of \newtool with existing tools by employing the dataset collected in \S~\ref{sec:evaluation:RQ1}. 
All tools construct their own TPL databases using the same dataset (6,819 TPL versions).} All compared tools choose similarity comparison method to find in-app TPLs, thus, the detection time mainly depends on the number of in-app TPLs and the number of TPL features in the database. 
The detection time is the period cost for finding all TPL-Vs in a test app. 
Note that the detection time does not include the database construction time.

\noindent\textbf{Result:} Table~\ref{tbl:detection_time} shows the comparison result of detection time. 
We present four metrics (i.e., Q1, mean, median, Q3) to evaluate the efficiency of each tool. 
We can see that the efficiency of \newtool also outperforms the state-of-the-art tools
{(66.24s per app on average)}. The second one is LibScout, and the average detection time is about 83s. LibID and LibPecker are relatively time-consuming; the average detection time could reach about 16.56h and 4.5h per app.
%

\begin{table}[]
\centering
\caption{Comparison Results of Detection Time {(per app)}. }
\scalebox{0.94}{
\begin{tabular}{lccccc}
\toprule[0.4mm]
\textbf{Tool} & \textbf{ATVHunter} & \textbf{LibID} & \textbf{LibScout} & \textbf{OSSPoLICE} & \textbf{LibPecker} \\ 
\midrule[0.4mm]
\textbf{Q1} & 15.92s & 51.43s & 30s & 33.48s & 12168s \\
\textbf{Mean} & 66.24s & 59616s & 83s & 2052.34s & 16396s \\
\textbf{Median} & 47.78s & 9286s & 64s & 80.42s & 16632s \\
\textbf{Q3} & 90.30s & 38300s & 100s & 226.60s & 23292s \\ \hline
\end{tabular}
}
\label{tbl:detection_time}
\end{table}

\newtool is more efficient than others because our method only needs to directly search to find the matching pairs in most situations, which can dramatically decrease the detection time.
\newtool employs a two-stage identification method (i.e., filter the potential TPLs first and identify the exact TPL with its specific version) to find the matched libraries from the database, which does not need to directly compare with the whole database using fine-grained features and largely reduces the comparison time and the whole detection time.  
In contrast, in the similarity feature comparison stage, LibScout needs to use the class dependency to filter some impossible pairs out, and this step is also time-consuming. Besides, LibScout regards the code of the host app as one of the candidate TPLs, which also costs extra time. 
OSSPoLICS exploits the fuzzy method signature (the same feature of LibScout)~\cite{libscout2016ccs} as the TPL code feature and function centroid~\cite{chen14ICSE} as the version code feature. The feature granularity of OSSPoLICE is much finer than that of LibScout, thus, the computational complexity of OSSPoLICE is also greater than that of LibScout. Besides, calculating centroid is heavy in terms of runtime overhead and computing resources consumption, especially for the third element (loop depth) in the centroid. The time complexity is $O((n+e)(c+1))$ and the space complexity is $O(n+e)$ to find all the loops, where there are $n$ nodes, $e$ edges and $c$ elementary circles in the graph. For LibPecker, if it tries to find a similar class, it needs to compare three times while our method only needs to compare once. Besides, LibPecker also needs to compare the package hierarchy structure and then 
{calculates} the similarity score, which also adds extra time. LibID chooses finer granularity features to identify TPLs, the class dependency analysis, CFG construction and class matching are also time-consuming. 

\vspace{1mm}
\noindent\fbox{
	\parbox{0.95\linewidth}{
		\textbf{Conclusion:} 
Compared with other tools, \newtool can identify exact TPL-Vs with high efficiency and it takes less time for TPL detection on the ground-truth TPL database. 
	}}
	
\subsection{RQ3: Obfuscation-resilient Capability}
\label{sec:evaluation:RQ4:obfuscation}
The obfuscation-resilient capability is an important index to measure the performance of a TPL detection tool since obfuscation techniques can discount the detection performance.


\noindent\textbf{Experimental Setup.}  
To evaluate the obfuscation-resilient capability of \newtool regarding different obfuscation techniques, we select 100 apps from 
the public dataset~\cite{orlis_benchmark} including multiple categories, and use a popular obfuscation tool, Dasho~\cite{DashO}, to obfuscate these apks with four widely-used obfuscation techniques (i.e., renaming obfuscation, control flow randomization, package flattening and dead code removal). 
Obfuscation is a time-consuming task and requires the obfuscation tool to analyze the code logic in order to conduct the obfuscation. It took us about half a month to obfuscate all of apps.
Finally, we get one group (100 apps) of the original apps and four groups ($100 \times 4$) of the obfuscated apps. 
Based on these groups of apps, we compare \newtool with other tools in terms of the detection rate ($\frac{|TP|}{|GT|}$) at version-level.

\begin{table}[]
 \small{\caption{Comparison on Code Obfuscation Techniques}}
\scalebox{0.8}{
\begin{tabular}{cccccc}
\toprule[0.4mm]
\multirow{2}{*}{\textbf{Tool}} & \multirow{2}{*}{\textbf{\begin{tabular}[c]{@{}c@{}}No\\ Obfuscation\end{tabular}}} & \multicolumn{4}{c}{\textbf{Obfuscation}} \\ \cline{3-6} 
 &  & \textbf{Renaming} & \textbf{CFR} & \textbf{PKG FLT} & \textbf{Code RMV} \\ 
\midrule[0.3mm]
\textbf{ATVHunter} & \textbf{99.26\%} & \textbf{99.26\%} & \textbf{90.13\%} & \textbf{99.26\%} & 75.57\% \\
\textbf{LibID} & 12.93\% & 12.93\% & 0.03\% & 1.58\% & 2.49\% \\
\textbf{LibScout} & 88.75\% & 88.75\% & 18.24\% & 17.69\% & 17.69\% \\
\textbf{OSSPoLICE} & 85.62\% & 85.62\% & 23.04\% & 39.52\% & 48.86\% \\
\textbf{LibPecker} & 98.79\% & 98.79\% & 86.63\% & 73.56\% & \textbf{79.28\%} \\ 
\bottomrule
\end{tabular}
}

  \begin{center}\small
	 \color{black}\textit{Renaming: renaming obfuscation; CFR: Control Flow Randomization; PKG FLT: Package Flattening; Code RMV: Dead Code Removal}
\end{center}
\vspace{-2ex}
\label{tbl:obfuscation}
\end{table}


\noindent\textbf{Result:} 
The detection results are presented in Table III,
the second column is the detection rate of each tool on apps without obfuscation. We can see ATVHunter achieves the highest detection rate (99.26\%), followed by LibPecker (98.79\%). 
Besides, it can be found that the detection rate of LibID is only 12.93\%, which has a big gap with the result in RQ1. 
We found the main cause of this gap is due to the inability of decompilation component dex2jar used by LibID. Many apps in this dataset cannot be decompiled successfully by dex2jar because of TPL compatibility issues, type errors and anti-decompilation settings, hence LibID cannot generate the in-app TPL signature, leading to the low detection rate.
%

{As for the capability of tools on obfuscated apps, we can see that all tools are resilient to renaming obfuscation since the detection rate of all tools on renaming apps is the same as the apps without obfuscation.
Our \newtool is less affected by all of these code obfuscation techniques. 
Code removal has the greatest impact on \newtool,
detection rate dropped by about 24\%. The detection rate on apps with other obfuscation techniques remains over 90\%, demonstrating the capability of \newtool towards commonly-used code obfuscation techniques. Moreover, we can find the recall of apps are obfuscated by package flattening is the same with the apps without obfuscation, it shows that our method is completely resilient to package flattening. In contrast, apart from the renaming obfuscation, the detection rate of other tools has been affected by obfuscations to varying degrees. 
Especially for LibScout, the performance has dropped by more than 70\%. LibScout can only correctly identify 17.69\% of in-app TPLs that are obfuscated by package flattening or dead code removal, and 18.24\% of in-app TPLs with control flow randomization. Except \newtool, LibPecker achieves better performance.} 

%
As for the control flow randomization (CFR), LibScout and OSSPoLICE use the fuzzy method signature as code features that keep the syntax information but do not remain semantic information; thus, it is difficult to defend against CFR. Besides, OSSPoLICE employs CFG centroid~\cite{chen14ICSE} as the version-level code feature. The CFG centroid is a three-dimensional vector, and each dimension indicates the in-degree, out-degree and loop count, respectively. The CFG centroid is sensitive to CFG structure modification; hence the detection rate of OSSPoLICE has dropped a lot regarding apps with CFR.
LibPecker and LibID show a good resiliency to CFR because both of them select the class dependencies as the code features that
{would not} be changed easily by CFR. \newtool extracts CFG as our coarse-grained feature and opcode in the basic block of CFG as the fine-grained feature. We keep the semantic information and remove the operands so our method is resilient to identifier renaming. 
We split the opcode sequence into small pieces and exploit fuzzy hash generate the code feature, although the dead code removal obfuscation and control flow obfuscation techniques can affect a part of code features, our strategy effectively reduces the interference, making the detection rate decline slightly. 

Regarding the package flattening technique, existing tools more or less depend on package structure to generate TPL signatures, without a doubt, which will affect their performance. More specifically,
LibScout depends on package structure/name to split TPLs. Firstly, many TPLs belong to the same group that may have the same package name. It is difficult to split these TPLs correctly if they belong to the same group. Secondly, the package flattening technique can easily change the package hierarchy structure or even remove the whole package tree, resulting in that LibScout will generate incorrect TPL signatures or cannot generate signatures for TPLs without package structures. OSSPoLICE is built on LibScout hence OSSPoLICE inherits the limitations of LibScout. LibPecker assumes the package structure is preserved during obfuscation but 
{it does not always hold true for real-world apps.}
This strong assumption directly restricts it to achieve better performance. 
In contrast,\newtool uses the class dependency relation to split different TPL candidates (on the basis of high cohesion and low coupling among different TPLs), which completely does not depend on the package structure, thus, \newtool is resilient to package flattening/renaming.

As for dead code removal, this obfuscation technique will delete some code that is not invoked by host apps, leading the code features of in-app TPLs are different from the original TPLs. This obfuscation can affect all TPL detection tools. LibPecker chooses class dependency as the code feature that keeps the method call relationship while we adopt CFG as code feature that do not include the method dependency. Our method may include methods and classes without invocations.
The signature of LibPecker stores more semantic information than that of us so that LibPecker achieves better performance in dead code removal.

\vspace{2mm}
\noindent\fbox{
	\parbox{0.98\linewidth}{
\textbf{Conclusion:} 
 \newtool offers better resiliency to code obfuscation than existing tools, especially for identifier renaming, package flattening, and control flow randomization.
}}


\section{Large-Scale Analysis}
\label{sec:measure}
{By leveraging \newtool,} we {further} conducted a large scale study on Google Play apps to reveal the threats of vulnerable TPL-Vs in the real world. 

\noindent \textbf{Dataset Collection.}
We collected commercial Android apps from Google Play based on the number of installations. For each installation range, we crawled the latest versions of apps from Aug. 2019 to Feb. 2020 for this large-scale experiment. We only consider popular apps whose installation ranges from 10,000 to 5 billion, because the vulnerabilities in apps with large installations can affect more devices and users. Note that the number of apps in each installation range is unequal; in general, the number of apps with higher installations usually is relatively {smaller}.
We finally collected 104,446 apps across 33 different categories as the study subjects. From our preliminary study on these apps, we found  72\% of them (73,110/104,446) use TPLs to facilitate their development. We thus focus on the 73,110 apps to conduct the following analysis.


\vspace{-1ex}
\subsection{Vulnerable TPL Landscape}
\vspace{-0.7mm}
\label{sec:large-scale:vulTPL_understanding}

Before conducting the impact analysis of vulnerable TPLs, we first present some essential information about these vulnerable TPL-Vs to let readers have a clear understanding about the threats in TPLs. 
We use CVSS v3.0 security metrics~\cite{cvss} to indicate the severity (i.e., low, medium, high, and critical) of vulnerabilities. The score greater than 7.0 means the vulnerability with high and critical severity, which accounts for 21.35\% of all the vulnerabilities in our dataset. These severe vulnerabilities usually involve remote code execution, sensitive data leakage, Server-side request forgery (SSRF) attack, etc.
Even worse, we find \textbf{74.95\%} of these vulnerable TPLs are widely-used by other TPLs.
For example, the library ``org.scala-lang:scala-library'' with a severe security risk ($CVSS$ = 9.8) that allows local users to write arbitrary class files, has been used 24,112 times by other TPLs, and most of vulnerable versions of this TPL have been used more than 2,000 times.
Without a doubt, such cases expand the spread of vulnerabilities and add more security risks to app users. These severe vulnerabilities usually involve remote code execution, sensitive data leakage~\cite{chen2019ausera,chen2018mobile}, malicious code or SQL injection, bypass certificates/authentication, etc. These behaviors definitely bring unpredictable
risks to users' privacy and property security. 
We found that most of these vulnerable TPLs belong to utility, accounting for 98.7\%. 

\vspace{-0.7mm}
\subsection{Impact Analysis of Vulnerable TPLs}
\vspace{-0.7mm}

In our dataset, we find that about 12.37\% (9,050/73,110) of apps include TPL-Vs, involving 53,337 known vulnerabilities and 7,480 security bugs from open-source TPLs. The known vulnerabilities are from 166 different vulnerable TPLs with corresponding 10,362 versions and the security bugs are from 27 vulnerable TPLs with 284 different versions.  
These vulnerable apps use a total of 58,330 TPLs and approximately 18.2\% of them are vulnerable ones.
Among the 9,050 vulnerable apps, 329 apps (37.5\%) with TPLs contain both vulnerabilities and security bugs. There are 778 apps containing the TPLs with security bugs and each app contains about 2.45 security bugs in their TPLs. 
Furthermore, we also find many education and financial apps use the popular UI library ``PrimeFaces''~\cite{prime} that include sever vulnerability (CVE-2017-1000486). Primefaces 5.x is vulnerable to a weak encryption flaw resulting in remote code execution. For more analysis result, you can refer to our website~\cite{atvhunter}.

\vspace{-0.5mm}
\subsection{Lessons Learned}
\vspace{-0.7mm}

Based on our analysis, we found many apps include vulnerable TPLs leading to privacy leakage and financial loss. However, developers seem unaware of the security risks of TPLs. 
We explore the reasons from the following points:

\noindent\textbf{For TPL developers,} according to our result in \S~\ref{sec:large-scale:vulTPL_understanding}, the reuse rate of vulnerable TPLs is pretty high ($>75\%$). Many TPL developers also develop their own TPLs based on existing ones, especially popular ones, but seem seldom to check the used components for any known vulnerabilities. 
{Even worse, we find 210,727 TPLs use vulnerable TPL versions, indicating many TPL developers may be unaware of tracking these vulnerability fix solutions in these open-source products. Although some TPL developers have patched the vulnerabilities in later versions, many affected apps still use the old versions with vulnerabilities, which indirectly expands the threats of the vulnerabilities in TPLs. 
The lack of centralized control of these open-source TPLs also poses attack surfaces for hackers. }

{\noindent\textbf{For app developers,}
we reported some TPL versions with severe vulnerabilities to the corresponding app developers via emails.
We wrote 50 emails to these app developers or companies and received 5 replies in 2 months. Based on their feedback, we find 1) most of the developers only care about the functionalities provided by the TPLs and are unaware of the security problems in these TPLs. {In fact, it is reasonable since one is unlikely to analyze all the used libraries before using them, which eliminates the convenience of using these components or libraries. However, based on our analysis, some commonly-used TPLs contain severe vulnerabilities, we suggest that app developers should be aware of vulnerabilities in TPLs and \newtool could be helpful for them to detect vulnerable TPL versions.
}
%
2) Some app developers or companies do not know how to conduct security detection of these imported TPLs. They also hope ``our team can help them conduct the security assessment of the used TPLs or tell them the specific analysis processes.''
3) Some app developers did not know that some vulnerable TPLs have been updated or patched and they still used these old TPL versions.
%
Even if they noticed the upgraded versions, some of them are reluctant to change the old ones due to the extra cost. They said that ``If a TPL adds many new functions, they have to spend much time understanding these new features and change too much of their own code. Thus, they prefer to keep old TPL-Vs.''
}

\noindent\textbf{For app markets,} 
we found that many app markets do not have such a security assessment mechanism to 
warn developers about the potential security risks in their apps. As far as we know, only
Google provides a service named App Security Improvement (ASI) program that provides tips to help app developers of Google Play to improve the security of their apps. Previous research~\cite{yasumatsu2019CODASPY} reported that vulnerabilities listed on ASI program could draw more attention from developers. However, the vulnerabilities reported by ASI program are limited due to the lack of a comprehensive vulnerability database and such a vulnerable TPL detection tool, like \newtool.

\section{Discussion}
\label{sec:discussion}
\noindent \textbf{{Limitations.}}
(1) If the Java code of several versions is the same, \newtool would provide several candidates instead of a specific one, leading to some false positives.
%
(2) \newtool may eliminate some TPLs due to mistakenly regarding them as part of the primary module if such TPLs are imported into the package structure of the host app, thus causing some false negatives. 
(3) We only focus on the Java libraries and do not consider the native libraries. In fact, the native library is also an essential part in Android apps and the vulnerabilities inside would cause more severe consequences. Detecting vulnerable native libraries is left for our future work. 
%
(4)\newtool adopts static analysis to find the TPLs, therefore, we may miss some libraries are loaded in dynamic methods. Besides, some TPLs have some dynamic behaviors, such as refection, dynamic class loading. Our approach may miss some dynamic features and affect our detection performance.
(5) We crawled about 3 million TPLs from maven to build our feature database. Although this database is large and comprehensive and it can guarantee the detection rate of \newtool, our method still have some limitations. The third-party libraries are constantly updating, which means \newtool cannot find these newly emerging TPLs. Thus, how to find these newly emerging TPLs and dynamically maintain our database will be our future work.

\noindent \textbf{Threats to Validity.} 
(1) The first threat comes from the similarity threshold, it is inevitable to induce some false negatives and false positives for some apps due to the minor difference between TPLs.
To minimize the threat, we selected the similarity threshold through a reasonable experimental design.
(2) Another threat comes from the analysis on only free apps. 
We believe that it is meaningful to study the vulnerable TPLs used by both free and paid apps, which is left for future work.
\vspace{-1.mm}
\section{Conclusion}
\label{sec:conclusion}

In this paper, we proposed \newtool, a TPL detection system which can precisely pinpoint the TPL version and find the vulnerable TPLs used by the apps. Evaluation results show that \newtool can effectively and efficiently find in-app TPLs and is resilient to the state-of-the-art obfuscation techniques. Meanwhile, we construct a comprehensive and large vulnerable TPL version database containing 224 security bugs 
and 1,180 CVEs. 
\newtool can find the vulnerable TPLs in apps and reveals the threat of vulnerable TPLs in apps, which can help improve the quality of apps and
has profound impact on the Android ecosystem.

\balance

\section{Acknowledgment}

We thank the anonymous reviewers for their helpful comments. This work is partly supported by the National Research Foundation, Prime Ministers Office, Singapore under its National Cybersecurity R\&D Program (Award No. NRF2018NCR-NCR005-0001), the Singapore National Research Foundation under NCR Award Number NRF2018NCR-NSOE003-0001, NRF Investigatorship NRFI06-2020-0022, the Singapore National Research Foundation under NCR Award Number
NRF2018NCR-NSOE004-0001, the Hong Kong PhD Fellowship Scheme and Hong Kong RGC Projects (No. 152223/17E,152239/18E, CityU C1008-16G).

\clearpage
\small
\bibliographystyle{IEEEtran}
\bibliography{LibsurveyBiB}

\end{document}